# Acceptor-based silicon quantum computing

B. Golding and M.I. Dykman

Department of Physics & Astronomy

Michigan State University

E. Lansing, MI 48824-2230

**Abstract**

A solid-state quantum computer with dipolar coupling between qubits is proposed. The qubits are formed by the low-lying states of an isolated acceptor in silicon. The system has the scalability inherent to spin-based solid state systems, but the spatial separation between the qubits is an order of magnitude larger. Despite strong dipolar inter-qubit coupling, the decoherence rate, as measured by electric dipolar echoes at an energy splitting of 1.5 GHz, is less than 1 kHz at low temperatures. For inter-acceptor distances of 100 nm and for modest microwave field amplitudes (50 V/cm) the clock frequency of the quantum computer is 0.1 GHz, which yields a quality factor of $10^5$. This paper describes ideas for detection and operation of the quantum computer, and examines limitations imposed by noise sources.

The concept of a solid-state quantum computer (QC) based on silicon has many attractive features. In comparison to a QC based on atoms, ions or liquid-state NMR, the anticipated integration with existing silicon device fabrication and nanoscale technology should provide overwhelming advantages. The major challenge is to identify a scalable system that has a sufficiently long decoherence time, can be controlled with high precision, and whose quantum state can be measured. Examples of systems based on silicon proposed to date include localized electron spins in semiconductor quantum dots(*1-4*); donor electron spins in Si/Ge heterostructures(*5*); a zero nuclear spin $^{28}$Si matrix with nuclear spin qubits on $^{31}$P donors interacting via overlap of localized electron wave functions(*6, 7*); chains of $^{29}$Si nuclei(*8*); and deep impurity states coupled with an optical field (*9*).

Spin-based systems are natural candidates for quantum computers, as they take advantage of extremely long spin coherence times. However, their implementation is complicated by a number of issues yet to be resolved. For nuclear spins of donors, for example, inter-donor separations should be less than 20 nm(*6, 7*). Execution of gate operations is relatively slow. Detection using a single-electron transistor appears feasible but the close donor separation makes fabrication highly demanding. These arguments suggest that there is a need to consider new alternatives for quantum computing in the condensed phase.

In this paper we describe a scalable solid-state quantum computer with "spinless" qubits. The states of the qubit are the low-lying states of a substitutional group III acceptor in silicon. The proposed system has many features in common with the NMR systems in liquids and semiconductors. It has the scalability inherent to spin-based solid state systems. However, the required spatial separation between the qubits is an *order of magnitude larger* than in the impurity-spin system(*6*), which greatly simplifies fabrication. This is made possible by strong dipolar coupling between the qubits. Measurements described here indicate a decoherence rate of order 1 kHz for this system for qubits with level separation 1-2 GHz. The frequency of gate operations is controlled by the interaction between the qubits and by the Rabi frequency in a resonant microwave field. For inter-acceptor distances of 100 nm and for moderate microwave power, the clock frequency of the quantum computer is 0.1 GHz. This implies a quality factor of $10^5$, thus satisfying the criterion for quantum error correction. Different states of the proposed qubits have different charge distributions, which makes it possible to detect the state of a qubit with a capacitive measurement. Alternatively, it is possible to take advantage of the extremely narrow linewidths of acceptor-bound excitons in isotopically pure silicon for near-infrared detection (*10*).

The isolated neutral shallow acceptor reflects the electronic structure of the $p_{3/2}$ valence band states at **k**=0 in silicon. As shown in Fig. 1, external electric and strain fields lift the four-fold degeneracy of the ground state, and the resulting Kramers doublets can be further split in a magnetic field. The static energy levels are readily tuned into the GHz range by strains $\sim 10^6$ Pa and by electric fields $\leq 10^3 - 10^4 \, \text{Vcm}^{-1}$. The two states of the qubit are the ground and first excited state of the acceptor. Qubit transitions are excited by resonant microwave electric field pulses in a low-Q microwave cavity.

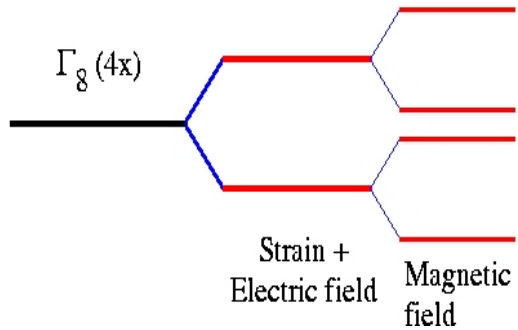

Fig. 1. Splitting of the shallow acceptor energy levels in Si by external fields. The ground state has a 4-fold degeneracy that is split by external strain and by an electric field. It is additionally split by a magnetic field.

Experiments conducted in the low microwave region between 0.5 and 10 GHz, provide the framework for understanding the couplings and decoherence mechanisms of acceptors in Si. The decoherence times at cryogenic temperatures have been measured with electric dipolar echoes from shallow acceptors (B, In, and Ga) in lightly-doped silicon crystals. These are the electric-field analog of spin echoes, with matrix elements that depend on the induced electric dipole moment connecting the two qubit states. The echoes are generated by placing the doped silicon samples into the electric field region of a re-entrant cavity tuned near 1.5 GHz. Analysis of the echoes provides a quantitative evaluation of the induced electric dipole, the longitudinal relaxation time $T_1$, decoherence time $T_2$, and the inter-acceptor coupling strength.

We first consider the longitudinal relaxation time which is governed by a single-phonon process below 1 K. The acceptor-phonon interaction Hamiltonian projected onto the lowest states has the form

$$H_{e-ph} = \sum_{\mathbf{q}j}\sum_{n,\alpha} V_{\mathbf{q}j\alpha n} s_\alpha^n (b_{\mathbf{q}j} + b_{-\mathbf{q}j}^+) \exp(i\mathbf{q}\mathbf{R}_n), \qquad (1)$$

where $s_\alpha^n$ ($\alpha = x, y, z$) are pseudo-spin operators which act in the two-state basis of the $n^{th}$ acceptor ($\mathbf{R}_n$ is the acceptor position) and $b_{\mathbf{q}j}, b_{\mathbf{q}j}^+$ are annihilation and creation operators of a phonon with the wave vector $\mathbf{q}$ and branch $j$. The matrix elements of electron-phonon coupling $V_{\mathbf{q}j\alpha n}$ vary as $\propto q^{1/2}$ in the deformation potential approximation and depend on the symmetry of the field that causes the splitting of the acceptor levels. The relaxation rate for one-phonon emission is explicitly given by(*11*)

$$T_1^{-1} = \frac{1}{\pi \hbar \rho c_t^5} MD^2 \omega^3 \coth\left(\frac{\hbar \omega}{2k_B T}\right) \qquad (2)$$

where $D$ is a deformation potential, $M$ is a geometrical factor of order unity, $c_t$ is the transverse acoustic sound velocity, and $\rho$ is the mass density.

The decay of a three-pulse electric field sequence in a sample of Si:In doped at 9 x 10$^{15}$ cm$^{-3}$ is shown in Fig. 2. The echo decay is highly non-exponential and persists for a time $T_1 \sim 1$ ms. The long lifetime results from the relatively low density of states of phonons in Si at energies $\hbar\omega \sim 1.5$ GHz despite the large values, ~3 eV, of the deformation potential $D$.

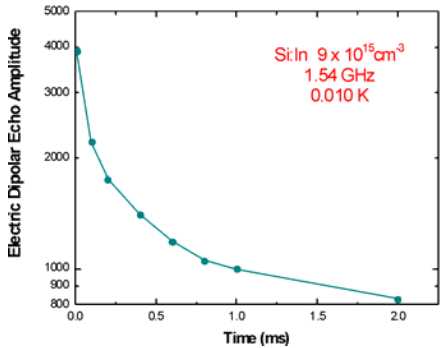

Fig. 2. Decay of 3-pulse dipolar electric echo from In acceptors in Si. The echo is generated with a $\frac{\pi}{2}, \frac{\pi}{2}, \frac{\pi}{2}$ sequence. Its decay is a measure of the lifetime of the excited state. In the absence of acceptor-acceptor interactions the decay is exponential and yields $T_1$ directly.

In the bulk samples described here, the mean acceptor-acceptor separation is 30 nm, and strong inter-acceptor coupling leads to a foreshortened T$_2$ of order 10 μs. The inter-qubit coupling is $\sum_{\mathbf{q}\alpha} V_{\mathbf{q}j\alpha n} V^*_{\mathbf{q}j\alpha' m} \exp[i\mathbf{q}(\mathbf{R}_n - \mathbf{R}_m)]/\hbar\omega_j(q)$, where $\omega_j(q)$ is the phonon frequency. The relevant phonon frequencies are $\omega_{mn} = \omega_j(q = 2\pi/d_{mn}) \sim 2\pi c_t/d_{mn}$. If they exceed the transition frequency $\omega$ significantly, as is the case here, the interqubit coupling exceeds the decay rate by a factor $\propto (\omega_{mn}/\omega)^3$. The detailed form of the coupling depends on the matrix elements $V_{\mathbf{q}j\alpha n}$ which can be calculated explicitly for specific electronic configurations of acceptors.

The non-exponential echo decay in Fig. 2 provides evidence for the long-range interaction between acceptors. The decay is explained by two-state elastic dipoles that interact with each other via phonons, with an interaction of the form of $D^{nm} s_z^m s_z^n / d_{nm}^3$. Here, $D^{mn}$ is the effective coupling constant and $d_{nm}$ is the dipole-dipole distance. For the acceptor doping densities in Fig. 2, the energy splittings of neighboring acceptors differ significantly. Therefore, we disregard all terms in the coupling $\propto s_\alpha^m s_\beta^n$ with $\alpha, \beta \neq z$. Spectral diffusion of the excitation across the broad line takes place by small energy shifts caused by flipping of thermally excited dipoles(*12, 13*). The echo decay contains a diffusion kernel that leads to a non-exponential temporal decay in Fig. 2 as well as in two-pulse echoes. The deformation potential of the elastic dipoles obtained from analysis of these echo decays is approximately 3 eV. In the dilute limit, and in the absence of thermally excited dipoles, these processes will be absent, provided $\hbar\omega > k_B T$. Therefore, we expect that $T_2 \sim T_1 \sim$ 1 ms.

Based on the experimental data, we conclude that the elastic dipolar interaction between neighboring qubits leads to a two-qubit gate frequency of ~0.1 GHz at 100 nm.

The experiment shows that the split acceptor ground state is accompanied by the appearance of a transition electric dipole moment. Direct measurements, using the condition for maximizing the amplitude of the two-pulse electric dipolar echo, give an orientationally-averaged value ~ 2-3 Debye for B and In acceptors. Thus, the Rabi frequency in a resonant microwave field with amplitude 50 V/cm is also 0.1 GHz, which gives the overall clock frequency of 0.1 GHz.

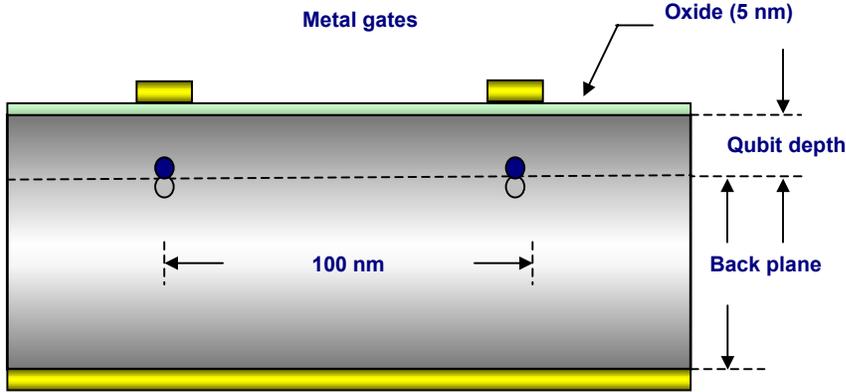

Fig. 3. Schematic diagram of the acceptor-based computer. The level splitting can be controlled by voltages applied to individual gates, shown here separated by 100 nm.

The general scheme of the acceptor-based quantum computer is shown in Fig. 3. Using a focused ion beam and a mask, acceptors are implanted within a 10 nm lateral window at a depth ~10 nm beneath the interface. They form a regular structure, with acceptor-acceptor distances of order 100 nm. Each acceptor is addressed by a single, individual gate that imposes a static electric field, and tunes the level spacing via the Stark effect.

The different electronic states of the acceptor in an electric field are associated with different charge distributions. This allows one to measure the state populations by capacitance changes induced on the electrode and detected with an electrometer positioned adjacent to the gate. The radius of the localized electron state is much less than the distance $h$ from the acceptor to the gate. Therefore a change of the state of the qubit results in the change of the dc potential on the gate $\delta V \sim \hat{P}_Z / \varepsilon h^2$, where $\hat{P}_Z$ is the dipole moment associated with difference in the charge distributions in the ground and excited states of the acceptor (see below), and $\varepsilon$ is the dielectric constant of silicon. For readout, the static dipole moment can be enhanced momentarily by applying a strong electric field from the electrode. For $\hat{P}_Z \sim 10$ Debye and $h \sim 10$nm, the potential $\delta V$ is of order 1 mV. The >100 nm gate separation makes SET integration relatively straightforward. Its use here has much in common with measuring charge transfer in a spin-based scheme(14).

Luminescence linewidths of acceptor-bound excitons in isotopically purified silicon have been measured to be < 0.005 cm$^{-1}$ (10). These are remarkably narrow lines, an order of magnitude narrower than the splitting of the qubit states. Therefore luminescence can be resonantly excited in the near infrared from one or from the other acceptor state. By detecting an emitted photon it will be possible to determine whether the state was occupied prior to excitation.

The interqubit distance of order 100 nm leads to straightforward fabrication by implantation of Ga acceptors with a focused ion beam. Implant damage will need to be minimized by suitable thermal annealing. The strain state of the near-surface region can be used to advantage in manipulating the distribution of qubit static splittings and the orientations of qubit states.

Solid state quantum computers with dipolar qubits pose a number of theoretical problems, including both the development of new quantum algorithms and investigation of quantum decoherence and entanglement. A distinctive feature of such computers, compared to systems of the liquid-NMR type (*15*) is that, even though for both electric and magnetic dipoles the interaction between the qubits may not be turned on and off, the transition frequencies of electric dipoles can be varied by individually applied electric fields via a Stark shift. Therefore a multi-qubit system can be controlled using a microwave source with only one or perhaps a few frequencies, by tuning targeted qubits in resonance with radiation. This allows one to implement single-qubit gates and to refocus the qubit-qubit interaction (*16*).

Both CNOT and SWAP two-qubit operations can be implemented using a combination of dc pulses on the control electrodes and microwave pulses. If we make a reasonable assumption that, during a quantum computation, the electrons may only be found in the ground and excited states $|0\rangle \equiv |\downarrow\rangle$ and $|1\rangle \equiv |\uparrow\rangle$ or their superposition, the electron Hamiltonian in terms of the pseudo-spin-½ operators $s_\alpha^n$, becomes

$$H = H_{con} + H_{unc}, \quad H_{con} = \sum_n \left[\hbar \omega_n(t) s_z^n + F_n(t) s_x^n\right],$$

$$H_{unc} = \frac{1}{2} \sum_{n \neq m} \left[A_{nm} s_z^n s_z^m + B_{nm} s_-^n s_+^m\right]$$

(3)

The term $H_{con}$ describes the one-qubit part, which can be controlled, with $\hbar\omega_n = E_{n1} - E_{n0}$ being the electrode-potential dependent energy of the 0→1 transition of the *n*th qubit, and $F_n(t)$ the scaled microwave field. The frequencies $\omega_n$ can be individually and independently varied over a range at least of order $\pm 0.1\langle\omega_n\rangle$. This allows tuning targeted qubits in and out of resonance with the external radiation and with each other. The radiation frequency will be in the dynamical range of variation of all frequencies $\omega_n$.

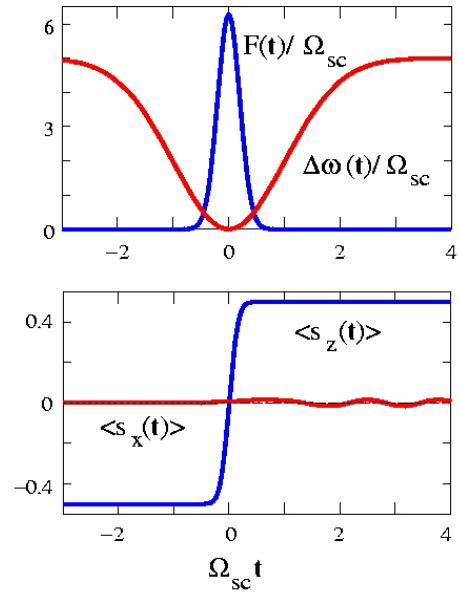

Fig.4. Excitation of a qubit, initially in the ground state $|0\rangle$ (with $\langle 0|s_z|0\rangle = -1/2$), by simultaneously applied pulses of microwave radiation $F(t)$ and frequency detuning $\Delta\omega(t) = \omega(t) - \omega_F$, where $\omega(t)$ is the controlled qubit frequency and $\omega_F$ is the microwave frequency. The scaling frequency $\Omega_{sc}$ is chosen so that $\int F(t)dt = \pi\hbar$. By varying amplitudes and widths of the pulses $\omega(t)$ and $F(t)$ one can drive the qubit into an arbitrary state.

The term $H_{unc}$ describes the qubit-qubit interaction. This interaction is dipolar for nearest neighbors, and falls off at large inter-qubit distances for the phonon-mediated coupling between acceptors. The term $H_{unc}$ cannot be controlled. We note that the Hamiltonian (3) describes also operation of quantum computers based on Josephson boxes(*17*) and electrons on helium(*18*). It applies to liquid-NMR systems as well, but the terms $\propto B_{nm}$ are rapidly oscillating and therefore effectively small, as the frequencies of different qubits differ significantly and may not be independently changed.

The frequencies $\omega_n/2\pi$ are of order 1-5 GHz, and therefore for temperatures 10 mK the qubits of the proposed quantum computer approach the ground state over the relaxation time $\geq 10^{-3}$ s. Initially the qubits are tuned away from resonance with the microwave radiation and with each other. A pulse of microwave radiation makes an exponentially small effect on a nonresonant qubit, whereas a resonant qubit can be efficiently excited. A single-qubit operation consists of applying a radiation pulse and simultaneously capacitively tuning a targeted qubit in resonance with the radiation, as shown in Fig. 4. This provides a complete and highly selective set of single-qubit gates. It is straightforward to calculate optimal pulse shapes, incorporating constraints consistent with transmission lines of the controlling electrodes.

The interaction Hamiltonian $H_{unc}$ (3) makes it possible to perform two-qubit operations. The first term in this Hamiltonian is the same as the Hamiltonian used to perform CNOT in the liquid-state NMR quantum computers. However, the second term in (3) allows us to perform a new type of quantum operation, a swap of excitation between qubits. This is done by tuning the targeted neighboring qubits in resonance with each other, as shown in Fig. 5. Depending on how the frequency difference $\Delta_{12}(t) = \omega_1(t) - \omega_2(t)$ is changed with time, the initial excitation can be redistributed between the qubits in the desired way. For example, the system can go from the state $|01>$ to $\alpha|01> + \beta|10>$, with arbitrary $\alpha, \beta$ ($|\alpha|^2 + |\beta|^2 = 1$).

The operation of excitation swap can be performed with exponential efficiency, because it is similar to Landau-Zener tunneling. It provides a powerful tool for generating

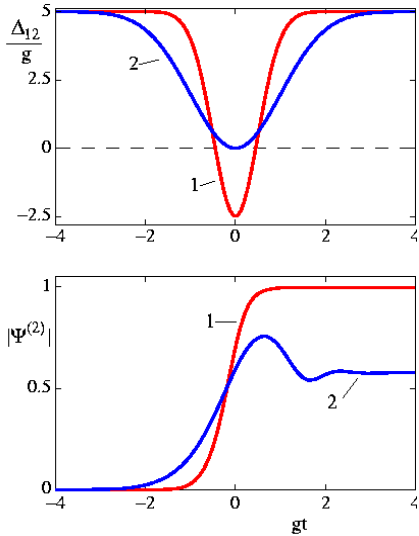

Fig.5. Excitation transfer between the qubits 1 and 2, which are initially ($t \to -\infty$) in the excited and ground states, respectively; $\psi^{(2)}$ is the wave function of the excited state localized on the qubit 2. The difference $\Delta_{12} = \omega_1 - \omega_2$ of the transition frequencies of the qubits is changed in time by applying a Gaussian pulse to the electrode potential. The qubits are in resonance with each other for $\Delta_{12} = 0$. The final amplitude of the excited state of the qubit 2 depends on the shape and amplitude of the applied pulse. The scaling frequency is $g = |B_{12}|/2\hbar$.

entangled states. For example, making the transformation

$$|01\rangle \xrightarrow{U_{12}} (|01\rangle + |10\rangle)/\sqrt{2}$$

with the excitation swap operator $U_{12}$, and then applying NOT to the qubit 2, we will obtain the Bell state $(|00\rangle + |11\rangle)/\sqrt{2}$. Unitary gates $U_{12}$ can be constructed to perform the transformations $\alpha|01\rangle + \beta|10\rangle \to \alpha'|01\rangle + \beta'|10\rangle$.

An important part of the operation of a dipolar-coupled computer is recoupling (refocusing) by the control functions $F_n(t)$ and $\omega_n(t)$. Recoupling is used in NMR to selectively turn on/off the interaction between targeted qubits. Since the interaction between dipolar-coupled qubits is on all the time, the development of efficient recoupling methods is essential. It will be discussed elsewhere (*16*).

The most obvious sources of decoherence of acceptor-based qubits are inelastic scattering by phonons, which gives rise to the finite lifetime of the excited state $T_1$, and quasi-elastic (Raman) phonon scattering off acceptors, which gives rise to phase breaking. From the experimental data we have seen that the relaxation rate of acceptors is extremely small.

A specific type of error is related to so-called side-band absorption, in which a microwave-induced transition is accompanied by creation or annihilation of a phonon. This mechanism is important for many types of quantum computers(*19*). It leads to quantum errors during certain gate operations. However, these errors do not accumulate between operations. Analyzing this mechanism with the small-polaron transformation, we find, that because of the extremely low density of phonon states at low frequencies of the order of the QC clock frequency, the associated errors are extremely small for acceptor-based qubits.

Coupling to the electrode may also provide an important mechanism of relaxation of a confined electron (qubit). One can readily estimate the effect of noise of gate electrodes and the dissipation related to electron-hole density fluctuations in the electrodes. Fluctuations of the electrode potential modulate the inter-level distance via Stark effect and thus give rise to dephasing. In addition, an electron can make a transition between the states, with energy being transferred to an excitation in the electrode (for example, an electron-hole pair). The analysis is similar to that done for electrons on helium (20). It is simplified by the fact that the size of the wave function of the qubit (the Bohr radius of the acceptor $a_B$) is small compared to the distance to the electrode $h$. Then the interaction of a qubit with the electrode can be described in the dipole approximation as

$$H_{gate} = -\sum_\alpha \delta \mathbf{E} \cdot \hat{\mathbf{P}}_\alpha s_\alpha, \tag{4}$$

where $\delta \mathbf{E}$ is the fluctuating part of the field on the electron from charge density fluctuations in the electrode, $\hat{\mathbf{P}}$ is the operator of the dipole moment of the qubit, and $s_\alpha$ is the $\alpha$ th component of the pseudo-spin operator for the electron system (we use Greek letters to enumerate the components of the pseudo-spin operator; they should be distinguished from the spatial components of the field and the dipole moment, which we enumerate by capital X,Y,Z). The vectors $\hat{\mathbf{P}}_\alpha$ are related to the diagonal and off-diagonal matrix elements of the operator of the electron dipole moment $\hat{\mathbf{P}}$ by the expressions $\hat{\mathbf{P}}_z = \langle 1|\hat{\mathbf{P}}|1\rangle - \langle 0|\hat{\mathbf{P}}|0\rangle$, $\hat{\mathbf{P}}_x + i\hat{\mathbf{P}}_y = \langle 0|\hat{\mathbf{P}}|1\rangle$.

Electron relaxation parameters can be expressed in terms of the correlation function of the fluctuating electric field from the electrode

$$Q_{IJ}(\omega) = \int_0^\infty dt \exp(i\omega t) < \delta E_I(t)\, \delta E_J(0) >, \tag{5}$$

that is, in terms of the electrode noise spectrum. We will assume that the function $Q_{IJ}(\omega)$ is smooth in the frequency ranges of interest for dissipation effects, i.e. for either $\omega < kT/\hbar$ or $\omega \sim \omega_0$, where $\omega_0$ is the qubit frequency.

Field-induced time variation of the phase difference of the wave functions |1> and |0> is equal to $\delta\phi(t) - \delta\phi(t') = -\int_{t'}^{t} dt_1 [\delta \mathbf{E}(t_1) \cdot \hat{\mathbf{P}}_z(t_1)]/\hbar$. Classical (thermal or quasi-thermal) field fluctuations give rise to phase diffusion on times that greatly exceed the correlation time of the field $\delta \mathbf{E}$, which we assume to be short, $< \hbar/kT$. The coefficient of phase diffusion is equal to the dephasing rate, which from (5) has the form

$$\Gamma_\phi = \mathrm{Re} \sum (\hat{\mathbf{P}}_z)_J (\hat{\mathbf{P}}_z)_I \, Q_{IJ}(0)/\hbar^2 \qquad (6)$$

If the electrode noise spectrum has peaks at low frequencies $\omega \leq \Gamma_\phi$, or the noise is non-Gaussian, time evolution of the phase difference becomes more complicated than simple diffusion, and decay of the qubit becomes nonexponential in time. Although the analysis has to be modified in this case, it is still convenient to relate decoherence of qubits to the fluctuating field of the electrode $\delta \mathbf{E}$.

Decay rate of the qubit $\Gamma_d$ is given by the probability of a field-induced transition from the excited to the ground states of the qubit, $|1\rangle \to |0\rangle$. This probability is determined, in turn, by quantum fluctuations of the field $\delta \mathbf{E}$ at the qubit frequency $\omega_0$. From (5),

$$\Gamma_d = \mathrm{Re} \sum (\hat{\mathbf{P}}_+)_J (\hat{\mathbf{P}}_-)_I \, Q_{JI}(\omega_0)/\hbar^2 \ . \qquad (7)$$

Here we assumed that decay is due to spontaneous emission only, i.e. that there are no induced processes with energy transfer of the order of the qubit level separation $\hbar \omega_0$.

To estimate relaxation parameters for qubits based on acceptors we will assume that the controlling electrode is a conducting sphere of a small radius $r_{el}$ placed at a distance $h$ from the qubit. This model corresponds to the case where the size of the electrode tip is small compared to $h$. For low frequencies the surface of the sphere is equipotential. Then the fluctuating field of the electrode is simply related to its fluctuating potential $\delta V$.

Much of the low-frequency noise is due to voltage fluctuations from an external lead attached to the electrode, which has resistance $R_{ext}$ and is held at temperature $T_{ext}$ that may exceed the temperature of silicon. The noise can be found from Nyquist's theorem and gives the dephasing rate $\Gamma_\phi \sim kT R_{ext} |\hat{P}_z|^2 r_{el}^2 / \varepsilon^2 \hbar^2 h^4$. For $R_{ext} = 50$ Ohm, $T_{ext} = 10$ K, $\hat{P}_z = 1$ Debye, and $r_{el} = h = 10$ nm, this rate is $\sim 0.5 \times 10^3\, \mathrm{s}^{-1}$. Much of the impedance of the measurement instrumention is reactive and does not give rise to Johnson noise.

In contrast to low-frequency noise, high-frequency voltage fluctuations from sources outside the cryostat can be filtered out. Much of high-frequency quantum fluctuations of the electric field that affect a qubit come from the underlying microelectrode itself. They depend on the interrelation between the electron relaxation time $\tau_{el}$ in the electrode and $\omega_0$. If $\tau_{el}\omega_0 \ll 1$, the electrode conductivity does not display dispersion up to frequencies $\omega_0$. An order-of-magnitude estimate of the decay rate can be made by assuming that the controlling electrode is a lead, with resistance $R_{el}$, attached to a sphere, and it is the total charge on the sphere that experiences quantum fluctuations due to the quantum Nyquist noise in the lead. This gives an estimate $\Gamma_d \sim \omega_0 R_{el} |\hat{P}_+|^2 r_{el}^2 / \varepsilon^2 \hbar h^4$. For $R_{el} = 1\Omega$ the estimated rate becomes $< 0.1$ s$^{-1}$. We do not expect that this result will change dramatically when the real geometry of the electrode is taken into account. The major mechanism of decay is therefore the coupling to phonons discussed above.

The present ideas can be readily extended to dipolar qubits based on laterally quantized electron states in single-electron broad quantum wells in GaAs heterostructures. Our estimates show that, with lateral confinement of 0.1-0.3 μm, the decoherence rate for the corresponding qubits will be small compared to the transition frequency and to the frequency of interqubit interaction, provided the interqubit distance is of order 1 μm.

Thanks to D. Lidar, Z.S. Gribnikov, P.M. Platzman, R. Hull, and W. H. Haemmerle for valuable discussions and for assistance with the experiments. We acknowledge support of NSF-ITER and the Institute for Quantum Sciences at Michigan State University.